\documentclass[11pt]{article}

\usepackage{arxiv}

\renewcommand{\shorttitle}{OpenPM: Auditable Point-in-Time Evaluation for LLM Portfolio-Management Agents}

\usepackage[round]{natbib}        
\usepackage{tgpagella}            
\usepackage{latexsym}
\usepackage{amssymb}
\usepackage{amsmath}
\usepackage{booktabs}
\usepackage{colortbl}
\definecolor{grouprow}{gray}{0.92}
\usepackage{tikz}
\usepackage{enumitem}
\usetikzlibrary{positioning,arrows.meta,fit,backgrounds}
\usepackage{pgfplots}
\pgfplotsset{compat=1.18}

\usepackage[T1]{fontenc}

\usepackage[utf8]{inputenc}

\usepackage{microtype}

\usepackage{inconsolata}

\usepackage{graphicx}

\usepackage{url}
\usepackage[hidelinks]{hyperref}

\title{OpenPM: Auditable Point-in-Time Evaluation for\\ LLM Portfolio-Management Agents}

\author{
  \textbf{Xinying Cai\textsuperscript{1,}}\thanks{Equal contribution.},
  \textbf{Minghao Guo\textsuperscript{1,}}\footnotemark[\value{footnote}],
  \textbf{Jiahe Liu\textsuperscript{2}},
  \textbf{Jiaojiao Han\textsuperscript{3}},
  \textbf{Bangwei Guo\textsuperscript{1}},
  \textbf{Yitao Long\textsuperscript{4}},
\\
  \textbf{Yuxuan Chen\textsuperscript{5}},
  \textbf{Bohan Wu\textsuperscript{6}},
  \textbf{Dimitris N. Metaxas\textsuperscript{1}},
  \textbf{Raymond Li\textsuperscript{7,}}\thanks{Corresponding author.}
\\[5pt]
  {\small
  \textsuperscript{1}Rutgers University,
  \textsuperscript{2}Technical University of Denmark,
  \textsuperscript{3}New Jersey Institute of Technology,
\\
  \textsuperscript{4}New York University,
  \textsuperscript{5}Columbia University,
  \textsuperscript{6}University of Illinois Urbana-Champaign,
\\
  \textsuperscript{7}University of British Columbia}
}

\begin{document}
\maketitle
\begin{abstract}

Large language models are increasingly used to read markets, assess risk, and allocate capital. However, reported results for LLM trading agents can be inflated by look-ahead leakage, optimistic execution, and risk mandates that are described but not enforced. We present \textbf{OpenPM}, an auditable point-in-time evaluation framework for LLM portfolio-management agents. 
In OpenPM, an agent manages a \$1M long-only book over the S\&P 500 universe using market data at five-minute intervals. Every record visible to the agent must be available at the decision time. Natural-language risk mandates are converted into typed constraints and enforced on the executed portfolio. Each run produces audit artifacts, including a contamination certificate, a cost-sensitivity curve, and a constraint-adherence report.
We also build a reference agent named the tiered allocator, where typed analysts score candidates, a constructor LLM proposes weights, and a deterministic critic guarantees feasibility. We isolate constructor behavior by capturing analyst evidence once and replaying it across constructor models.
In our short-window case study, stronger constructors show modest and model-dependent gains over equal weighting on the same pool, but analyst quality matters more than constructor choice, and turnover is the main cost driver. All returns are upper bounds on a single frozen window without market impact, not validated alpha. OpenPM's code and data are available at \href{https://github.com/aslcai/OpenPM-Bench}{\nolinkurl{github.com/aslcai/OpenPM-Bench}} and \href{https://huggingface.co/datasets/aslcai/OpenPM-Bench}{\nolinkurl{huggingface.co/datasets/aslcai/OpenPM-Bench}}.

\end{abstract}

\section{Introduction}

Large language models are increasingly used in financial decision systems that read market context, assess risk, and allocate capital~\citep{yang2023fingpt,wu2023bloomberggpt, zhang2024ai}. For a practitioner, the key question is not whether an agent looks strong in an offline backtest, but whether that result would survive deployment. The agent must use only information that was available at the decision time, its returns must survive trading costs, and its final portfolio must respect the stated risk mandate. Small choices in data access, execution modeling, and constraint handling can create apparent skill that disappears in production~\citep{li2025deepfund,chen2025stockbench}. We treat credible evaluation as an engineering artifact, and build OpenPM as one.

We focus on three practical failures in LLM trading-agent evaluation that a \emph{benchmark}, not a better model, must control. \textbf{(i) Information leakage:} agent-visible records can include information that was not knowable at the decision time, such as features that use future observations or records stamped by event time rather than availability time~\citep{bailey2014pseudo,lopezdeprado2018advances}; this is closely related to NLP benchmark contamination~\citep{sainz2023nlp}. \textbf{(ii) Optimistic execution:} returns reported without spread, slippage, and turnover costs are upper bounds, not deployable estimates~\citep{chen2025stockbench}. \textbf{(iii) Unenforced mandates:} natural-language risk mandates, such as ``conservative'' or ``at most 20 names,'' are often used only for post-hoc scoring rather than enforced on the executed portfolio, so a model that ignores its mandate is never caught.
Although OpenPM is instantiated in portfolio management, these failures are not unique to trading. Trading is a useful stress test for agent evaluation~\citep{guo2026memeye,xu2026memgym} because it combines temporal grounding, multiple information sources, constrained action, noisy outcomes, and cost-sensitive execution. The same pattern appears in many deployed agents acting in open-ended environments~\citep{xu2026ael}: unavailable information can leak into the state, simulators can be too optimistic, and natural-language instructions can be treated as soft preferences rather than enforceable constraints.


We present \textbf{OpenPM}, which addresses these failures directly. An agent manages a single \$1M long-only book over an S\&P~500 universe, emitting dollar-denominated orders at an hourly, daily, or once-then-held cadence over 5-minute market-state observations. A single availability gate fixes information access: no agent-visible record enters a decision at time $T$ unless its availability timestamp is ${\le}\,T$, and future-looking labels are held out for scoring. Every run is scored not only on return, but also on \emph{whether} it used unavailable information, \emph{how} sensitive it is to execution costs, and \emph{whether} it obeyed its mandate. As a reference agent, we build a \emph{tiered allocator} (\S\ref{sec:allocator}), where typed analyst LLMs score candidates, a constructor LLM proposes weights, and a deterministic in-loop risk critic projects the proposal onto the feasible set. Thus, a deterministic projection, not the model, guarantees that the constraints hold.


Our contributions are:
\begin{itemize}[leftmargin=*]
\itemsep0em
\item \textbf{A point-in-time portfolio benchmark.}
We formalize availability-gated evaluation for portfolio agents over an S\&P~500 universe, with label-free price, liquidity, regime, event, news, and filings state under a single data contract and execution-aware labels held out for scoring.

\item \textbf{An accountability-first evaluation protocol.}
Each run produces a per-model contamination certificate, a transaction-cost-sensitivity curve, a constraint-adherence report, per-name and sector attribution, and a backtest-bias checklist, so every reported number is tied to the conditions that produced it.

\item \textbf{A reference allocator and diagnostic case study.}
We build a tiered allocator separating analyst evidence, constructor decisions, deterministic risk projection, and execution. On frozen evidence it gives four findings: construction gains are modest and model-dependent, equal-weighting a strong candidate set is a hard baseline, analyst quality matters more than constructor choice, and turnover drives cost. The deterministic critic is necessary because it enforces feasibility rather than trusting the LLM. Returns are no-market-impact upper bounds, not validated alpha.

\end{itemize}

We are deliberate about scope: following rigor-forward trading benchmarks~\citep{li2025deepfund,chen2025stockbench,yu2025livetradebench}, we make \emph{no validated-alpha claim}: fills are priced off single-venue IEX TOPS (not NBBO), the window is short, and prompts were iterated on overlapping data, so we report relative ordering rather than absolute return \emph{levels}. All results are tied to a frozen dataset fingerprint on a fixed commit.

\section{Related Work}

Table~\ref{tab:related} situates OpenPM against representative related work; it contrasts design, not performance, with deliberately conservative labels.

\begin{table}[t]
\centering
\small
\setlength{\tabcolsep}{6pt}
\resizebox{\textwidth}{!}{%
\begin{tabular}{@{}llcccc@{}}
\toprule
\textbf{Benchmark / agent} & \textbf{Task / universe} & \textbf{Cadence} & \shortstack{\textbf{NL mandate }\\ \textbf{to portfolio}} & \shortstack{\textbf{Leakage}\\\textbf{control}} & \shortstack{\textbf{Cost-aware}\\\textbf{labels}} \\
\midrule
FinBen~\citep{xie2024finben} & Text / QA & --- & No & None & No \\
FinMem, TradingAgents~\citep{yu2023finmem,xiao2024tradingagents} & Single-stock agent & Daily & Verbal & None & Limited \\
FinCon~\citep{yu2024fincon} & Portfolio, 3 assets & Daily & Verbal (CVaR) & None & Limited \\
InvestorBench~\citep{li2025investorbench} & Mostly single-stock & Daily & No & Partial & Partial \\
QuantAgent~\citep{xiong2025quantagent} & Directional, per-instr. & 1h / 4h & No & Sampled & No \\
MarketSenseAI~\citep{fatouros2025marketsenseai} & Stock-pick $\to$ portfolio & Monthly & No & Live post-cutoff & No \\
StockBench~\citep{chen2025stockbench} & Portfolio, DJIA-20 & Daily & No & Contam.-free & Partial \\
DeepFund~\citep{li2025deepfund} & Portfolio, 5 tickers & Daily & No & Post-cutoff & Partial \\
LiveTradeBench~\citep{yu2025livetradebench} & Portfolio, 15 names & Daily, live & No & Live forward & Partial \\
PortBench~\citep{zhao2026portbench} & Portfolio, 6 asset cls. & Monthly & Post-hoc (PAS) & PiT replay & Partial \\
\midrule
\textbf{OpenPM (ours)} & \textbf{PIT portfolio, S\&P~500} & \textbf{intraday\,--\,once} & \textbf{Enforced (critic)} & \textbf{Row-gate + cert.} & \textbf{Yes} \\
\bottomrule
\end{tabular}}
\caption{Design comparison with representative LLM trading benchmarks and agents. \textbf{NL mandate to portfolio} indicates whether natural-language investor constraints are enforced on the portfolio weights. \textbf{Leakage control} summarizes the benchmark's look-ahead, contamination, and point-in-time availability mechanism. OpenPM is the only one combining an S\&P~500-scale equity universe, sub-daily observations, row-level availability gating, critic-enforced mandate compliance, and cost-aware labels.}
\label{tab:related}
\end{table}

\paragraph{Financial LLMs and trading agents.}
Domain models and holistic benchmarks~\citep{yang2023fingpt,wu2023bloomberggpt,xie2024finben} target language understanding, not point-in-time trading. Single-system trading agents~\citep{yu2023finmem,li2023tradinggpt,xiao2024tradingagents,yu2024fincon,fatouros2025marketsenseai,li2025hedgeagents,zhao2025alphaagents,xiong2025quantagent} add memory, role structure, and verbal risk discussion, but are agent systems rather than reusable benchmarks, decide daily or slower over a handful of assets, and never bind a natural-language mandate to the executed weights, often reporting large historical returns without leakage or cost controls (Table~\ref{tab:related}).

\paragraph{Rigor-forward agent benchmarks.}
The closest cohort foregrounds contamination~\citep{chen2025stockbench,li2025deepfund,yu2025livetradebench,li2025investorbench}. PortBench~\citep{zhao2026portbench}, closest to us, ingests natural-language investor profiles but scores mandate alignment post hoc rather than enforcing it, and finds 90\% of model--profile pairs fail to beat equal-weight, consistent with our finding that equal-weight is hard to beat. All operate over small universes with no row-level availability gate; OpenPM instead (Table~\ref{tab:related}) pairs an S\&P~500-scale, sub-daily-to-once portfolio task with a per-record availability gate, a per-run contamination certificate, and a deterministic in-loop critic that enforces the mandate on the weights, following methodology work on contamination~\citep{sainz2023nlp} and backtest overfitting~\citep{bailey2014pseudo,lopezdeprado2018advances,lo2002sharpe,harvey2016cross}.

\section{Benchmark Design}
\label{sec:design}

\subsection{Task}
OpenPM frames portfolio management as a sequence of point-in-time rebalancing decisions over a single shared book. The agent starts with \$1{,}000{,}000 cash and at each decision time $T$ receives a label-free cross-section as of $T$. It observes this cross-section every 5-minute bar but \emph{rebalances} on a configurable schedule (hourly, daily, or once), emitting a possibly empty list of dollar-denominated \texttt{buy}/\texttt{sell} orders; anything not mentioned is held. The book is long-only and unlevered, enforced structurally (buys clip to cash, sells to shares held), trading whole shares. Buys fill at the ask and sells at the bid, so the side-aware half-spread is charged at execution time, not assumed away. Universe, number of decisions, and scores are properties of a data release, not the task.

\subsection{Point-in-Time Data Environment}
The central invariant is that agent-visible information must be available no later than the decision time at which it is used. Every record carries an availability timestamp and enters the state at decision time $T$ only if \texttt{ts\_available}~${\le}~T$. This is distinct from event time and bar-end time: a quarterly filing is stamped by its EDGAR \texttt{acceptanceDateTime}, macro series by their FRED/ALFRED vintage, and bar-derived features become available only after the bar closes. Gating on availability is designed to prevent conditioning on information unavailable at $T$.

The gate is exercised by unit tests and re-checked at run time by the contamination certificate (\S\ref{subsec:eval}), designed to prevent look-ahead leakage and audited for the modes we test. The agent-visible state exposes six typed layers, each computable point-in-time: \emph{price/bar}, \emph{liquidity}, \emph{regime}, \emph{event}, \emph{news}, and \emph{filings}. The fields in each layer and their availability stamps are given in App.~\ref{sec:schema}. Agents do not observe raw price levels; price information is provided through derived and normalized features, which reduces direct price-level memorization risk.

\subsection{Natural-Language Risk Mandates}
\label{subsec:risk}
A run carries a risk mandate. A preset word maps deterministically to typed \texttt{RiskConstraints} (\emph{conservative}: ${\le}5\%$ per name, ${\le}30$ names, $10\%$ cash; \emph{balanced}: ${\le}10\%$, ${\le}20$ names; \emph{aggressive}: ${\le}25\%$, ${\le}10$ names). Free-form text (``a cautious retiree'', an investor persona and value profile~\citep{du2025twinvoice,du2025simvbg}) is parsed by an LLM via forced tool-use into the same typed envelope with a confidence; a low-confidence or contradictory mandate is flagged and defaulted to balanced. Parsing is configuration-time, outside the leakage surface. Unlike verbal risk discussion in prior agents, the constraints are typed once and then enforced.

\subsection{Labels and Evaluation Targets}
Labels are evaluation-only and future-looking, barred from the agent at the file and column level. We score each book on its executable (cost-aware) return: fills clear at the quoted side, charging the side-aware half-spread (no slippage or market impact). Corporate-action adjustment makes the book total-return capable (dividends reinvested as cash).

\begin{figure}[!t]
\centering
\resizebox{\textwidth}{!}{%
\begin{tikzpicture}[
  font=\footnotesize,
  node distance=2mm and 2mm,
  llm/.style={draw, rounded corners=2pt, fill=blue!10, minimum height=9mm, minimum width=19mm, align=center, inner sep=1pt},
  det/.style={draw, fill=black!6, minimum height=9mm, minimum width=19mm, align=center, inner sep=1pt},
  result/.style={draw, fill=green!10, rounded corners=2pt, minimum height=9mm, minimum width=19mm, align=center, inner sep=1pt},
  audit/.style={draw, dashed, fill=yellow!15, minimum height=6.5mm, minimum width=19mm, align=center, inner sep=1pt, font=\scriptsize},
  banner/.style={draw, fill=red!8, minimum height=5.5mm, align=center, inner sep=2pt, font=\scriptsize\itshape},
  arrow/.style={-{Latex[length=1.6mm]}, thick},
  link/.style={-{Latex[length=1mm]}, dashed, gray},
]
\node[banner, text width=166mm] (gate) {Point-in-time availability gate: a record enters the agent-visible state at decision time $T$ only when $\texttt{ts\_available} \le T$};

\node[det, below=3.5mm of gate.west, anchor=north west] (s1) {NL mandate\\$\to$ typed};
\node[det, right=of s1] (s2) {Candidate\\gate};
\node[llm, right=of s2] (s3) {Six LLM\\analysts};
\node[det, right=of s3] (s4) {Aggregate\\+ top-$K$};
\node[llm, right=of s4] (s5) {Constructor\\LLM};
\node[det, right=of s5] (s6) {Risk\\critic};
\node[det, right=of s6] (s7) {Fill\\simulator};
\node[result, right=of s7] (s8) {Equity\\curve};

\foreach \a/\b in {s1/s2, s2/s3, s3/s4, s4/s5, s5/s6, s6/s7, s7/s8}
  \draw[arrow] (\a) -- (\b);

\node[audit, below=5mm of s1] (a1) {RiskConstraints\\manifest};
\node[audit, below=5mm of s2] (a2) {Gate log};
\node[audit, below=5mm of s3] (a3) {Prompt \&\\score logs};
\node[audit, below=5mm of s4] (a4) {Score\\matrix};
\node[audit, below=5mm of s5] (a5) {Raw\\proposal};
\node[audit, below=5mm of s6] (a6) {Projection\\diff};
\node[audit, below=5mm of s7] (a7) {Fill ledger};
\node[audit, below=5mm of s8] (a8) {Contam.\ cert.\\Bias checklist};

\foreach \s/\a in {s1/a1, s2/a2, s3/a3, s4/a4, s5/a5, s6/a6, s7/a7, s8/a8}
  \draw[link] (\s.south) -- (\a.north);

\end{tikzpicture}}
\caption{The OpenPM evaluation pipeline. A point-in-time availability gate (top) governs what each stage may see; the 8 stages turn a natural-language mandate into an executed, audited portfolio, calling an LLM only at the \emph{analysts} and \emph{constructor} (the rest are deterministic) and emitting a versioned audit artifact (bottom row) at every step.}
\label{fig:arch}
\end{figure}
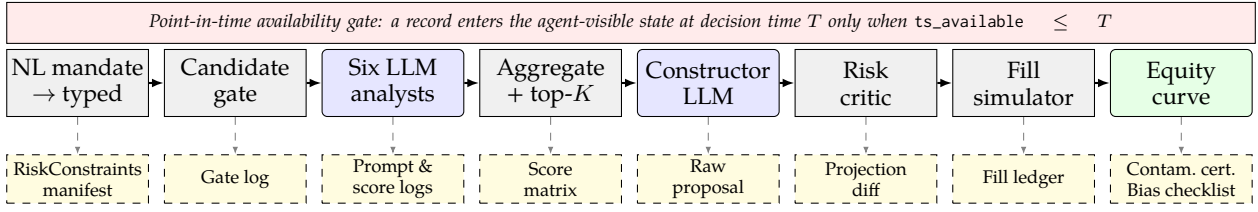

\section{The Tiered Allocator}
\label{sec:allocator}
The flagship agent is an eight-stage tiered allocator mapping the multi-analyst ``trading-firm'' template onto the leak-gated layers, with deterministic stages bracketing the LLM calls. The firm metaphor is not the contribution. It is saturated in prior agents~\citep{xiao2024tradingagents,yu2024fincon}; what is specific to OpenPM is its instantiation on point-in-time data: typed leak-gated analyst channels (including 8-K item codes as a structured, acceptance-time-gated feed, a leak-safe channel, not a validated signal source), a natural-language mandate compiled to typed constraints, and a deterministic in-loop critic that projects the weight vector onto them at decision time rather than scoring it afterward.

The architecture is shown in Figure~\ref{fig:arch}. A natural-language mandate is first compiled into typed constraints. A liquidity-only gate then bounds the cross-section to at most 50 tradable names. Six parallel cheap-model analysts independently score typed information channels---technical, regime, events, news, 8-K item codes, and BM25-retrieved 10-K/10-Q narrative ---with a score in $[-1,1]$ and a confidence value; routing such heterogeneous evidence to specialized analysts relates to LLM-based information routing~\citep{guo2026deepsieve}. Separately, a deterministic liquidity analyst emits per-name position caps. We aggregate analyst scores using a confidence-weighted average, $\mathrm{comp}(s){=}\sum_a w_a c_{a,s} x_{a,s}/\sum_a w_a c_{a,s}$, with uniform $w_a=1$ after an IC-weighted variant failed out of sample. A stronger constructor then proposes long-only weights via tool-use, seeing the per-analyst matrix and spreads but not raw prices or (by default) the composite or candidate order. The deterministic critic then projects the proposal onto the feasible set (long-only, top-$N$ (the name-count cap), per-name and liquidity clipping, cash/gross/turnover scaling), and a fill simulator marks the executed book at the quoted side. Each stage is designed to be independently ablatable and emits a versioned audit artifact; backbones are bring-your-own-key and the analyst and constructor tiers can differ.

\section{Experimental Setup}
\label{sec:setup}
\paragraph{Dataset.} We freeze a point-in-time S\&P~500 universe consisting of 508 names that are active on at least one date between 2026-02-19 and 2026-05-01. Each run records a content hash of the frozen data snapshot. The evaluation window spans 2026-03-02 to 2026-05-01, covering 44 trading days and 3{,}432 five-minute bars. At each bar, membership is gated point-in-time, so only names active in the index at that bar are eligible; this yields to 504 active names during evaluation. Within each session, no rebalancing happens in the first 60 minutes (12 five-minute bars) after the open; this post-open gate also gives the rolling features time to warm up. The data is 2026-dated, after every evaluated backbone's knowledge cutoff.


\paragraph{Constructor-isolation setup.}
Our central once-mode experiment isolates the constructor by holding analyst evidence fixed. We first run the six-analyst tier and freeze its output as an analyst capture: per-name scores and confidences over the liquidity-gated ${\le}50$-name pool. We create six captures from two analyst backbones (\texttt{DeepSeek-V3.2}, \texttt{gpt-5}) and three seeds, since MoE serving is not perfectly bit-reproducible even at temperature zero. Each capture is then replayed through five constructors (\texttt{gpt-5}, \texttt{claude-opus-4.7}, \texttt{DeepSeek-V3.2}, \texttt{qwen3-max}, and \texttt{gpt-4o-mini}) at $K\in\{30,40,50\}$, reusing the cached analyst outputs rather than recomputing them~\citep{lin2025cache}. The constructor observes the per-analyst confidence matrix and spreads, but not raw prices, the composite score, or the candidate order; candidates are shuffled before construction. This yields 90 LLM constructor cells on one decision date, plus 216 baseline cells. The daily sweep in \S\ref{subsec:daily} complements this controlled setting by rerunning each constructor over 44 daily rebalances at $K=50$ with fixed \texttt{DeepSeek-V3.2} analysts, reporting one point estimate per constructor.

\paragraph{Baselines.} Every baseline consumes the same frozen capture and runs the same backtest and critic with no LLM call. \textbf{Same-pool equal-weight} (EW) buys the top-$N$ names by composite score with equal weights, testing whether a constructor improves over simply splitting capital across the highest-ranked candidates. \textbf{Composite-ranking} sizes by inverse volatility over positive-composite names. \textbf{Random-N} equal-weights a seeded random draw (ten seeds). Market references are \textbf{SPY} and \textbf{always-flat}.

\paragraph{Evaluation protocol.}
\label{subsec:eval}
Each run reports equity-curve metrics, including total return, Sharpe, net-of-cost Sharpe, Sortino, maximum drawdown, turnover, and exposure. We also report a five-point cost-sensitivity curve at $\{0, 0.5, 1, 1.5, 2\}\times$ the default transaction cost, together with a binary \emph{beat-all-benchmarks} verdict indicating whether the run outperforms equal-weight, SPY, and always-flat. For accountability, every run emits three audit artifacts: a constraint-adherence report, a per-run contamination certificate with status PASS/FAIL/UNKNOWN, and a backtest-bias checklist. Per-name and per-sector P\&L are reconciled to the aggregate equity curve.

\section{Results}
We analyze two regimes: a controlled once-mode constructor-isolation grid (\S\ref{subsec:constructor}--\S\ref{subsec:adherence}) and a realistic daily rebalancing sweep (\S\ref{subsec:daily}). We report ordering and structure, not precise sizing (\S\ref{sec:limitations}).

\begin{table}[!t]
\centering
\footnotesize
\setlength{\tabcolsep}{3pt}
\resizebox{\textwidth}{!}{%
\begin{tabular}{@{}l *{8}{r} @{\hspace{1.8em}} *{8}{r}@{}}
\toprule
 & \multicolumn{8}{c}{\textbf{\texttt{DeepSeek-V3.2} analysts} ($K{=}50$)} & \multicolumn{8}{c}{\textbf{\texttt{gpt-5} analysts} ($K{=}50$)} \\
\cmidrule(lr){2-9}\cmidrule(lr){10-17}
\textbf{Constructor} & \textbf{Net} & \textbf{Shp} & \textbf{MDD} & \textbf{Turn} & \textbf{$\Delta$EW} & \textbf{Beat} & \textbf{N} & \textbf{mW} & \textbf{Net} & \textbf{Shp} & \textbf{MDD} & \textbf{Turn} & \textbf{$\Delta$EW} & \textbf{Beat} & \textbf{N} & \textbf{mW} \\
\midrule
\rowcolor{grouprow}\multicolumn{17}{@{}l}{\textbf{LLM constructors}}\\
\texttt{gpt-5}        & $\mathbf{7.0}$ & $\mathbf{1.73}$ & $8.4$ & $0.96$ & $\mathbf{-1.0}$ & $\mathbf{5/9}$ & $16.9$ & $8.7$ & $\mathbf{12.5}$ & $\mathbf{2.81}$ & $7.8$ & $0.96$ & $\mathbf{+5.3}$ & $\mathbf{9/9}$ & $14.6$ & $8.8$ \\
Opus-4.7              & $6.4$ & $1.61$ & $8.4$ & $1.00$ & $-1.6$ & $2/9$ & $19.6$ & $8.3$ & $10.8$ & $2.61$ & $7.6$ & $0.94$ & $+3.6$ & $\mathbf{9/9}$ & $18.9$ & $8.3$ \\
DeepSeek-V3.2         & $2.2$ & $0.64$ & $8.4$ & $1.01$ & $-5.8$ & $2/9$ & $16.3$ & $7.4$ & $11.1$ & $2.51$ & $7.8$ & $0.95$ & $+3.9$ & $8/9$ & $14.6$ & $7.6$ \\
Qwen3-Max             & $1.1$ & $0.37$ & $9.0$ & $0.97$ & $-6.9$ & $0/9$ & $11.6$ & $8.7$ & $8.3$ & $1.76$ & $8.7$ & $0.92$ & $+1.1$ & $8/9$ & $10.3$ & $9.3$ \\
\texttt{gpt-4o-mini}  & $0.1$ & $0.12$ & $\mathbf{7.4}$ & $0.91$ & $-7.9$ & $0/9$ & $9.6$ & $9.3$ & $5.1$ & $1.28$ & $\mathbf{7.3}$ & $0.92$ & $-2.1$ & $4/9$ & $10.0$ & $9.4$ \\
\addlinespace[2pt]
\rowcolor{grouprow}\multicolumn{17}{@{}l}{\textbf{No-LLM baselines} (same frozen capture)}\\
Same-pool EW          & $8.0$ & $2.03$ & $7.9$ & $1.01$ & ---    & ---   & $19.9$ & $5.0$ & $7.2$  & $1.85$ & $8.9$ & $1.01$ & ---    & ---   & $19.9$ & $5.0$ \\
Composite-rank        & $0.6$ & $0.31$ & $5.6$ & $0.96$ & $-7.4$ & $0/9$ & $19.9$ & $8.7$ & $0.9$  & $0.41$ & $5.8$ & $0.96$ & $-6.3$ & $0/9$ & $19.9$ & $8.3$ \\
Random-$N$ (mean)     & $0.7$ & $0.32$ & $7.6$ & $1.03$ & $-7.2$ & ---   & $19.9$ & $5.0$ & $0.7$  & $0.32$ & $7.6$ & $1.03$ & $-6.5$ & ---   & $19.9$ & $5.0$ \\
\addlinespace[2pt]
\rowcolor{grouprow}\multicolumn{17}{@{}l}{\textbf{Market references}}\\
SPY buy-and-hold      & $5.8$ & $2.01$ & $8.6$ & --- & --- & --- & --- & --- & \multicolumn{8}{c}{\emph{Identical under \texttt{gpt-5} analysts:}}\\
Always-flat (cash)    & $0.0$ & $0.00$ & $0.0$ & --- & --- & --- & --- & --- & \multicolumn{8}{c}{\emph{objective market constants, not analyst-dependent.}}\\
\bottomrule
\end{tabular}}
\caption{\textbf{Constructor comparison} (balanced mandate, once-then-hold 2026-03-02 to 2026-05-01, $K{=}50$, means over three seeds). Columns split by the \emph{analyst backbone} producing the byte-identical frozen capture (\texttt{DeepSeek-V3.2} vs.\ \texttt{gpt-5}). \textbf{$\Delta$EW}: active return vs.\ same-pool EW (pp); \textbf{Beat}: cells beating same-pool EW over $3$ seeds${\times}3$ $K$ (pooled over all $K$, whereas Net and $\Delta$EW are $K{=}50$, so a backbone can show $\Delta$EW${<}0$ yet Beat${>}0$ by clearing EW at smaller $K$); \textbf{mW}: max single-name weight\,\% ($10\%$ cap); \textbf{bold}: best LLM per column; Gross omitted ($\approx$\,Net). Full per-$K$ multi-metric grid in Table~\ref{tab:grid}. Returns are single-window, no-market-impact upper bounds (\S\ref{sec:limitations}); ordering, not validated alpha.}
\label{tab:main}
\end{table}

\subsection{Can an LLM construct a portfolio?}
\label{subsec:constructor}
Table~\ref{tab:main} is the benchmark's central question: holding the analyst evidence byte-identical, can an LLM constructor beat splitting the book evenly across the same candidates? A qualified yes that depends on signal quality. The constructors order \texttt{gpt-5} $>$ Opus-4.7 $>$ DeepSeek-V3.2 $>$ Qwen3-Max $>$ \texttt{gpt-4o-mini}, an ordering stable when pooled across $K\in\{30,40,50\}$ (Beat~EW $14/18$, $11/18$, $10/18$, $8/18$, $4/18$; Sharpe tracks returns). 

The key dependence is the analyst capture. On the \texttt{gpt-5} capture, where same-pool EW returns $7.2\%$, four of five constructors beat EW at $K{=}50$ (\texttt{gpt-5}:$+5.3$ points; DeepSeek:$+3.9$; Opus:$+3.6$; Qwen:$+1.1$), with only \texttt{gpt-4o-mini} trailing ($-2.1$). However, on the \texttt{DeepSeek-V3.2} capture, none of the constructors beat EW; the best, \texttt{gpt-5}, reaches $\Delta$EW$=-1.0$ point. The inverse-volatility composite-ranking baseline is much weaker ($0.6$--$0.9\%$ net return, $0/18$ Beat-EW), so same-pool EW is the hard naive baseline; the upstream analyst lever this exposes is quantified in \S\ref{subsec:analyst}.

Structurally, the strongest constructors add value mostly by re-weighting the EW roster rather than replacing it: their holdings overlap with EW is approximately $0.75$--$0.78$. The weakest constructor departs most from the roster and loses (\texttt{gpt-4o-mini}: $0.39$ overlap; details in App.~\ref{sec:additional}). Overall, construction adds value when the upstream signal is strong enough to act on; otherwise, the constructor cannot reliably improve over the hard EW baseline.

\subsection{The analyst tier is the larger lever}
\label{subsec:analyst}
The constructor is not where most of the value lives. Re-running the same constructors on different analyst captures separates the two tiers: moving from the \texttt{DeepSeek-V3.2} analyst capture to the \texttt{gpt-5} analyst capture raises constructor performance across models, while same-pool EW changes only modestly ($8.0\%\to7.2\%$) (Table~\ref{tab:main}). This indicates that the stronger analyst capture does not merely improve the candidate pool; it provides evidence that capable constructors can exploit through re-weighting. Conversely, when the analyst signal is weaker, constructors often fall below same-pool EW, showing that construction cannot reliably rescue poor upstream signal. The practical implication is to spend compute first on the analyst tier and then use the cheapest capable constructor---a heterogeneous allocation of compute across specialized LLM components that echoes budget-controllable multi-LLM routing~\citep{mei2025omnirouter}.

\subsection{Mandate compliance and the critic's role}
\label{subsec:adherence}


A portfolio benchmark should distinguish raw mandate adherence from post-critic feasibility. Under the balanced mandate (per-name cap $10\%$, at most $20$ names), the raw proposals from all five constructors across six captures satisfy both explicit caps: the largest single-name weight is exactly the $10.0\%$ cap in 24 of 90 cells and never exceeds it, and no constructor proposes more than $20$ names. Thus, in this setting, the mandate's explicit weight and name-count caps do not bind after construction.

The critic's active work is instead liquidity enforcement. The per-name liquidity hard-clip, which can be tighter than the mandate and is not directly visible to the model, binds in 37 of 90 cells, concentrated in weaker constructors (\texttt{Qwen}: $16/18$; \texttt{gpt-4o-mini}: $17/18$). We report these liquidity clips separately from mandate breaches, so they are not miscounted as failures to follow the natural-language mandate.

This result shows that many constructors can follow simple numeric mandates, but it does not justify assuming compliance in general. The critic remains on the critical path as a deterministic feasibility layer. It enforces both the user mandate and trading constraints that the model may not observe, while the adherence report makes any breach measurable under tighter mandates or other constructors.

\subsection{Cost and cadence: turnover, not spread, governs net return}
\label{subsec:daily}
\begin{table}[t]
\centering
\small
\setlength{\tabcolsep}{6pt}
\begin{tabular}{@{}lrrrrc@{}}
\toprule
\textbf{Constructor} & \textbf{Net\%} & \textbf{Gross\%} & \textbf{Shp} & \textbf{Turn} & \shortstack{\textbf{Beat}\\\textbf{all}} \\
\midrule
Opus-4.7        & $\mathbf{13.7}$ & $\mathbf{14.8}$ & $\mathbf{4.32}$ & $22.9$ & \checkmark \\
DeepSeek-V3.2   & $9.7$  & $10.7$ & $3.01$ & $25.8$ & \checkmark \\
\texttt{gpt-4o-mini} & $5.1$ & $6.8$ & $1.77$ & $41.4$ & --- \\
Qwen3-Max       & $4.8$  & $7.4$  & $1.58$ & $43.6$ & --- \\
\texttt{gpt-5}  & $4.2$  & $5.5$  & $1.35$ & $21.6$ & --- \\
\bottomrule
\end{tabular}
\caption{Daily-cadence sweep: 44 rebalances, 2026-03-02 to 2026-05-01, balanced mandate, $K{=}50$, \texttt{DeepSeek-V3.2} analysts; single runs (point estimates). ``Beat all'' = beats SPY ($5.8\%$) and cash, net of cost.}
\label{tab:daily}
\end{table}
Because the candidate gate screens to a liquidity-ranked pool, spread is not the bottleneck; turnover is. In once-then-hold mode, the book turns over only at construction, so net and gross returns are nearly identical (\texttt{Opus}: $10.77\%$ gross vs.\ $10.76\%$ net, about $2$ bps per dollar traded). In daily mode, however, costs are charged at each of 44 rebalances. \texttt{Opus} and \texttt{DeepSeek} still beat the market references net of cost, returning $13.7\%$ and $9.7\%$ versus SPY's $5.8\%$ and cash (Table~\ref{tab:daily}). But cumulative turnover now rises to $22$--$44$ times, so the high-churn constructors \texttt{gpt-4o-mini} ($41$) and Qwen ($44$) pay $1.7$--$2.5$ percentage points of cost drag and slip below SPY.
Low turnover alone is not sufficient. \texttt{gpt-5} has the lowest turnover ($22$) but only $4.2\%$ net return. The winning pattern is therefore to select well and trade sparingly. Figure~\ref{fig:turnover} (Appendix~\ref{sec:additional}) plots net return against turnover and highlights \texttt{gpt-5} as a low-turnover, low-return exception. These single-run point estimates suggest that daily-or-slower cadence is viable for disciplined constructors, while also showing that per-model rankings are regime- and date-sensitive. We therefore emphasize structural findings rather than a fixed model ranking.

\subsection{Leakage and bias accounting}
The contamination certificate is PASS for all five constructors, whose knowledge cutoffs precede the 2026 window (\texttt{gpt-5} ${\sim}2024$-09, \texttt{gpt-4o-mini} ${\sim}2024$-10, \texttt{Qwen3-Max} ${\sim}2025$-06, \texttt{DeepSeek-V3.2} ${\sim}2025$-07, \texttt{Opus-4.7} ${\sim}2026$-01, the thinnest margin); all runs are causality-, chronology-, and parity-clean. The bias checklist flags cost realism, market impact, and data-snooping every run.

\section{Conclusion}
OpenPM fixes information access with a uniform availability gate and scores every run for leakage, cost optimism, and mandate adherence alongside return. Interrogating its flagship allocator on byte-identical evidence, we find genuine but conditional LLM construction skill: under a strong analyst capture most constructors beat equal-weighting the same candidates, under a weak one none do, and the upstream analyst tier moves returns more than the constructor choice. Skill localizes upstream of construction. At daily cadence the strongest constructors beat every market benchmark net of cost, with turnover the main cost driver. These are single-window, no-market-impact upper bounds, not validated alpha---the kind of claim a point-in-time, audited benchmark is built to keep honest.

\section*{Limitations}
\label{sec:limitations}
We scope OpenPM deliberately and interpret its results as diagnostic evidence rather than deployable trading performance. \textbf{Simulation idealization.} As in any backtest, fills clear at the quoted side, and we do not model market impact or queue dynamics. Net returns are therefore upper bounds at non-trivial notional and should be read for ordering rather than as achievable returns. Quotes are single-venue IEX TOPS rather than consolidated NBBO; for the liquidity-screened names traded by the allocator, this makes realized spread costs conservative, but it does not close the market-impact gap. \textbf{Scope of the snapshot.} Our controlled comparison is a point-in-time snapshot over one S\&P~500 window, with robustness drawn from repeated analyst captures rather than independent market regimes. We therefore report relative ordering and point estimates, not calibrated effect sizes or statistical alpha~\citep{lo2002sharpe,card2020power,harvey2016cross}. The ordering in Table~\ref{tab:main} should be read as snapshot evidence, since a small number of strongly trending names can drive much of the spread between constructors. Broader generalization across decision dates, intraday cadences, markets, and backbones is future work. \textbf{Coverage.} The universe is U.S.\ large-cap equities, and signals such as options and intraday fundamentals are out of scope. \textbf{Leakage.} The availability gate and contamination certificate are designed to enforce point-in-time access across the modes we evaluate. As with any audit, they reduce and bound leakage risk rather than proving that every possible configuration is leakage-free~\citep{10918255} .

\section*{Ethical Considerations}
This work is a research benchmark, not investment advice or a deployable trading system; its returns are explicitly upper bounds under an optimistic cost model and must not be read as achievable performance. We use only publicly available market, filing (SEC EDGAR), macro (FRED), and news (GDELT) data, within their terms, and do not redistribute licensed raw market data. LLM backbones are accessed under a bring-your-own-key scheme that never stores or logs secrets. Deploying LLM allocators on real capital without realistic cost, impact, and risk controls could cause financial harm; although the backbones we tested respected their numeric mandate, compliance cannot be assumed for tighter mandates or other models, which is why we keep a deterministic enforcement layer on the critical path rather than trusting the model. We disclose that AI assistants were used for code and writing support; all claims were verified by the authors against the saved run artifacts.

\section*{Acknowledgments}

\bibliographystyle{acl_natbib}
\bibliography{custom}

\appendix

\section{Point-in-Time Feature Contract}
\label{sec:schema}
This appendix documents what the agent observes and, for each field, when that evidence becomes available. The point is not to inventory columns but to show that the leakage gate of \S\ref{sec:design} is enforced field by field: every layer carries an availability stamp, and a field enters the state at decision time $T$ only once that stamp is ${\le}\,T$. All counts below are read from the freeze manifest and the released parquet/NDJSON files, not from intermediate reports.

\paragraph{Composition.} The release is a single content-hashed freeze of the 508 point-in-time S\&P~500 members active over the collection span; bar-level membership gating leaves 504 names tradable across the 44-day evaluation window (2026-03-02 to 2026-05-01, 3{,}432 five-minute decision bars per name). After point-in-time gating and cold-start removal this yields 1{,}739{,}121 agent-visible decision records, drawn from five precomputed five-minute layers ($\approx$2.0M rows each) and three evidence sources: 1{,}462{,}952 entity-resolved GDELT articles, 210{,}801 filing-body chunks across 506 symbols, and 2{,}020 Finnhub consensus rows.

\paragraph{Layers and availability.} The agent-visible state is a JSON object per (symbol, decision time) built from six typed evidence channels, each independently computable as of the decision bar. Table~\ref{tab:schema} lists the channels, their provenance, and the availability stamp that gates each---the same gate enforced in \S\ref{sec:design}. Agents never observe raw price \emph{levels}: price information enters only through the derived and normalized \texttt{bar\_state} fields, which limits direct price-level memorization. Five channels are dense per-bar layers; the sixth, \emph{filings}, is a decision-time retrieval channel consumed by two analysts (\texttt{filings\_meta} over gated 8-K item-code counts and \texttt{filings\_body} over leak-gated BM25 retrieval of 10-K/10-Q passages), so it is listed separately. The full per-field list ships with the released data card.

\begin{table}[t]
\centering\footnotesize
\caption{Evidence channels and their point-in-time availability. Each row is a typed channel the agent observes as of decision time $T$; the \emph{availability stamp} is the gate enforced in \S\ref{sec:design}---a field enters the state only once its stamp is ${\le}\,T$. \emph{Fields} gives the per-layer count with examples; \emph{filings} is a decision-time retrieval channel rather than a dense per-bar layer. Liquidity uses IEX TOPS venue-only top-of-book, not consolidated NBBO; the full field list ships with the released data card.}
\label{tab:schema}
\setlength{\tabcolsep}{4pt}
\begin{tabular}{@{}>{\raggedright\arraybackslash}p{1.4cm} >{\raggedright\arraybackslash}p{3.5cm} >{\raggedright\arraybackslash}p{2.5cm} >{\raggedright\arraybackslash}p{4.0cm} >{\raggedright\arraybackslash}p{3.0cm}@{}}
\toprule
Layer & Captures & Source & Availability stamp (PIT gate) & Fields ($n$, example) \\
\midrule
Price/bar & intraday returns, volatility, VWAP distance, gaps, time-of-day & IEX TOPS 5-min bars & bar close; 60-min post-open warmup & 14 (\texttt{return\_5m}) \\
\addlinespace
Liquidity & spread, quoted depth, illiquidity flag & IEX TOPS top-of-book & bar close & 6 (\texttt{mean\_spread\_bps}) \\
\addlinespace
Regime & market/sector-relative return, beta, breadth, curve slope, risk-on & SPY \& sector ETFs; FRED/ALFRED & bar close; FRED daily vintage (backward as-of); 5-day warmup & 15 (\texttt{risk\_on\_score}) \\
\addlinespace
Event & earnings timing, EPS surprise, post-earnings drift & SEC 8-K Item 2.02; Finnhub & 8-K \texttt{acceptanceDateTime}; next-event forecast only if $\le$21 days & 11 (\texttt{pead\_active}) \\
\addlinespace
News & news flow, recency, headline sentiment & GDELT GEG (entity-resolved) & publish $+\,$30-min lag; salience $\ge0.05$ & 4 (\texttt{news\_flow\_zscore}) \\
\midrule
Filings (retrieval) & 8-K item-code counts; 10-K/10-Q body passages & SEC EDGAR & \texttt{acceptanceDateTime}; gate-then-BM25 at $T$ & \texttt{filings\_meta}, \texttt{filings\_body} \\
\bottomrule
\end{tabular}
\end{table}

\section{Evaluation Protocol and Metric Definitions}
\label{sec:metrics}
Every metric is produced by a single deterministic evaluator from the executed equity curve; we give the exact definitions so reported numbers are reproducible. A \emph{bar} is a five-minute regular-session decision bar. The equity curve $E_0,\dots,E_T$ is anchored to the opening NAV $E_0$ (pre-trade equity at bar~0, so every strategy starts from the same capital and pays its bar-0 entry cost), and per-bar simple returns are $r_t=E_t/E_{t-1}-1$. The risk-free rate is taken as zero and the annualization factor is $\sqrt{78\times252}$ (78 bars per trading day).

\paragraph{Executable (cost-aware) return.} Fills clear at the quoted side---buys at the ask, sells at the bid---and held positions are marked at the mid, so the side-aware half-spread is realized immediately as a mark-to-mid markdown. A fill of $q$ shares costs $c=q\,\lvert p_{\text{fill}}-\text{mid}\rvert$, i.e.\ $q(\text{ask}-\text{mid})$ for a buy and $q(\text{mid}-\text{bid})$ for a sell. The headline \emph{net return} is $R=E_T/E_0-1$, with cost already embedded in $E_T$; the \emph{gross} (frictionless) return adds the realized cost back, $R_g=(E_T+\textstyle\sum_t c_t)/E_0-1$ (first-order), and the \emph{cost drag} is $R_g-R$. Books are corporate-action adjusted with dividends reinvested as cash, so $R$ is a total return.

\paragraph{Risk-adjusted metrics.} With $\mu_r$ and $\sigma_r$ the sample mean and standard deviation (denominator $n{-}1$) of $\{r_t\}$, the annualized Sharpe is $(\mu_r/\sigma_r)\sqrt{78\times252}$ at zero risk-free rate. Sortino uses the downside deviation $\sigma_r^{-}$ computed over the negative bars only, $\mu_r/\sigma_r^{-}$ (per-bar, not annualized). Maximum drawdown is the largest peak-to-trough fractional drop of the anchored curve.

\paragraph{Turnover and cost sensitivity.} Turnover is the total gross dollar volume traded over the run (buys and sells) divided by mean equity, $\tau=\sum_t \text{executed\_usd}_t/\overline{E}$ (whole-window, not annualized). The churn-adjusted Sharpe applies a heuristic penalty, $\mathrm{Sharpe}/(1+\tau)$, and return per unit turnover is $R/\tau$. The cost-sensitivity curve reports $R(m)=(E_T+(1-m)\sum_t c_t)/E_0-1$ for $m\in\{0,0.5,1,1.5,2\}$, where $m{=}1$ is the actual run and $m{=}0$ is frictionless. Exposure is the mean per-bar invested fraction $\overline{\text{invested}_t/E_t}$; long-only makes net and gross exposure coincide.

\paragraph{Active and beat metrics.} Each LLM book is compared, on the \emph{same} frozen analyst capture, against four no-LLM baselines---same-pool equal-weight (top-$N$ by composite), composite-ranking (inverse volatility over positive-composite names), random-$N$ (ten seeds), and the market references SPY buy-and-hold and always-flat cash. The active return $\Delta$EW is the paired difference $R_{\text{constructor}}-R_{\text{EW}}$ at the same capture and $K$ (in points). \emph{Beat} counts the cells in which $R_{\text{constructor}}>R_{\text{EW}}$: Table~\ref{tab:main} pools all $18$ cells (6 captures $\times$ 3 values of $K$), whereas the per-$K$ grid (Table~\ref{tab:grid}) reports wins out of 6. The \emph{beat-all-benchmarks} verdict is true iff $R$ strictly exceeds same-pool equal-weight, SPY, and always-flat, all net of cost; the daily sweep (Table~\ref{tab:daily}) reports the SPY-and-cash form.

\section{Additional Results}
\label{sec:additional}
\textbf{Selection overlap.} Jaccard overlap of each constructor's holdings with same-pool equal-weight (\S\ref{subsec:constructor}) ranges from $0.78$ (Opus, the least departure) and $0.75$ (\texttt{gpt-5}) down to $0.39$ (\texttt{gpt-4o-mini}, the most); mean pairwise overlap across the five constructors is $0.56$.

\noindent\textbf{Full per-$K$ grid.} Table~\ref{tab:grid} reports every constructor across $K\in\{30,40,50\}$, pooled over the six analyst captures (two backbones $\times$ three seeds). Three patterns hold across $K$: (i)~the ordering is stable (\texttt{gpt-5} $>$ Opus-4.7 $>$ DeepSeek-V3.2 $>$ Qwen3-Max $>$ \texttt{gpt-4o-mini}); (ii)~the edge over same-pool equal-weight shrinks as $K$ grows---a larger pool pulls every constructor toward EW, so $\Delta$EW falls (\texttt{gpt-5} $+2.9\!\to\!+2.1$\,pp, DeepSeek-V3.2 $+1.1\!\to\!-1.0$\,pp) and Beat rates decline; and (iii)~only \texttt{gpt-5} and Opus-4.7 clear EW at half or more of the captures for every $K$, whereas \texttt{gpt-4o-mini} clears it in at most $2/6$ and the inverse-volatility ranking baseline never ($0/6$). The weaker constructors also crowd the $10\%$ cap (mW $9.0$--$9.3\%$) into ${\approx}10$ names, while the stronger ones spread across $N{\approx}15$--$19$ well below the cap.

\begin{table}[t]
\centering\footnotesize
\setlength{\tabcolsep}{4pt}
\begin{tabular}{@{}llrrrr@{}}
\toprule
\textbf{Constructor} & $K$ & \textbf{Net\%} & \textbf{Shp} & \textbf{$\Delta$EW} & \textbf{Beat} \\
\midrule
\texttt{gpt-5}       & 30 & $10.5$ & $2.59$ & $+2.9$ & $5/6$ \\
                     & 40 & $9.8$  & $2.41$ & $+2.2$ & $5/6$ \\
                     & 50 & $9.7$  & $2.27$ & $+2.1$ & $4/6$ \\
\addlinespace[2pt]
Opus-4.7             & 30 & $9.5$  & $2.32$ & $+1.9$ & $4/6$ \\
                     & 40 & $9.0$  & $2.23$ & $+1.4$ & $4/6$ \\
                     & 50 & $8.6$  & $2.11$ & $+1.0$ & $3/6$ \\
\addlinespace[2pt]
DeepSeek-V3.2        & 30 & $8.7$  & $1.99$ & $+1.1$ & $3/6$ \\
                     & 40 & $7.7$  & $1.82$ & $+0.1$ & $4/6$ \\
                     & 50 & $6.6$  & $1.57$ & $-1.0$ & $3/6$ \\
\addlinespace[2pt]
Qwen3-Max            & 30 & $7.7$  & $1.72$ & $+0.2$ & $3/6$ \\
                     & 40 & $6.8$  & $1.52$ & $-0.8$ & $3/6$ \\
                     & 50 & $4.7$  & $1.06$ & $-2.9$ & $2/6$ \\
\addlinespace[2pt]
\texttt{gpt-4o-mini} & 30 & $5.6$  & $1.37$ & $-2.0$ & $2/6$ \\
                     & 40 & $3.1$  & $0.82$ & $-4.5$ & $1/6$ \\
                     & 50 & $2.6$  & $0.70$ & $-5.0$ & $1/6$ \\
\midrule
equal-weight         & all& $7.6$  & $1.94$ & ---    & ---   \\
ranking              & 30 & $1.2$  & $0.52$ & $-6.4$ & $0/6$ \\
                     & 40 & $0.8$  & $0.36$ & $-6.8$ & $0/6$ \\
                     & 50 & $0.8$  & $0.36$ & $-6.8$ & $0/6$ \\
\bottomrule
\end{tabular}
\caption{Per-$K$ constructor grid, means over six analyst captures (\texttt{DeepSeek-V3.2} and \texttt{gpt-5} backbones $\times$ three seeds), once-then-hold 2026-03-02 to 2026-05-01. \textbf{Net\%}: total net return; \textbf{$\Delta$EW}: active return vs.\ same-pool equal-weight (pp); \textbf{Beat}: captures (of six) beating same-pool EW. \texttt{gemini-2.5-flash} excluded (degenerate all-cash). Single-window upper bounds, not validated alpha (\S\ref{sec:limitations}).}
\label{tab:grid}
\end{table}

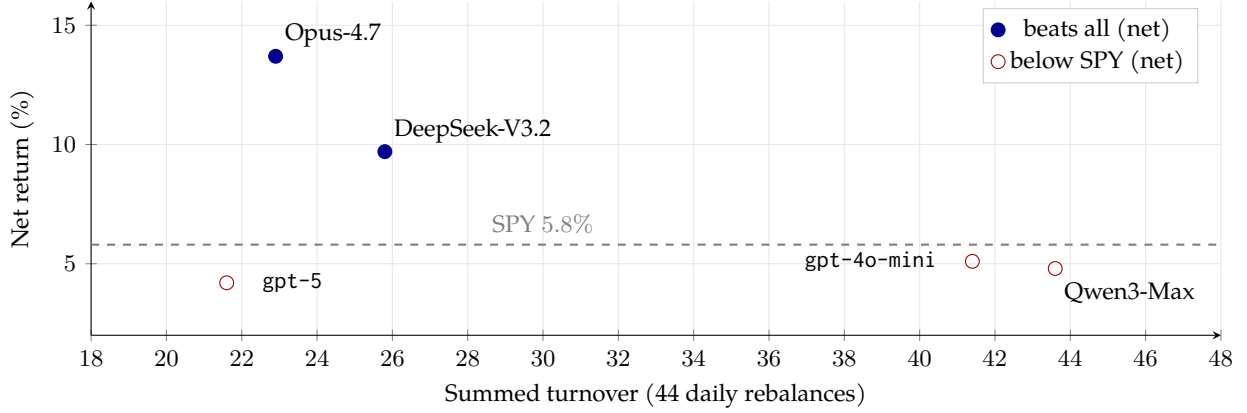
\begin{figure}[t]
\centering
\begin{tikzpicture}
\begin{axis}[
  width=\columnwidth,
  height=6cm,
  xlabel={Summed turnover (44 daily rebalances)},
  ylabel={Net return (\%)},
  xmin=18, xmax=48, ymin=2, ymax=16,
  grid=major, grid style={gray!18},
  tick align=outside, axis lines=left,
  label style={font=\small}, tick label style={font=\footnotesize},
  legend style={font=\footnotesize, at={(0.98,0.98)}, anchor=north east, draw=gray!40},
]
\addplot[dashed, gray, thick, domain=18:48, forget plot] {5.8};
\node[gray, font=\footnotesize, anchor=south] at (axis cs:30,5.9) {SPY $5.8\%$};
\addplot[only marks, mark=*, mark size=2.6pt, color=blue!55!black] coordinates {(22.9,13.7) (25.8,9.7)};
\addlegendentry{beats all (net)}
\addplot[only marks, mark=o, mark size=2.6pt, color=red!60!black] coordinates {(41.4,5.1) (43.6,4.8) (21.6,4.2)};
\addlegendentry{below SPY (net)}
\node[anchor=south west, font=\footnotesize] at (axis cs:22.9,13.7) {Opus-4.7};
\node[anchor=south west, font=\footnotesize] at (axis cs:25.8,9.7) {DeepSeek-V3.2};
\node[anchor=east, font=\footnotesize] at (axis cs:40.7,5.1) {\texttt{gpt-4o-mini}};
\node[anchor=north west, font=\footnotesize] at (axis cs:43.6,4.6) {Qwen3-Max};
\node[anchor=west, font=\footnotesize] at (axis cs:22.3,4.2) {\texttt{gpt-5}};
\end{axis}
\end{tikzpicture}
\caption{Net return vs.\ summed turnover at daily cadence (Table~\ref{tab:daily}, \texttt{DeepSeek-V3.2} analysts). The disciplined \texttt{Opus-4.7} and DeepSeek-V3.2 sit upper-left (low turnover, high net); the high-turnover \texttt{gpt-4o-mini} and Qwen sit lower-right, dragged below SPY by cost. \texttt{gpt-5} is the informative outlier: the \emph{lowest} turnover yet still below SPY, so trading sparingly is necessary but not sufficient---selection must also be good. Single runs (point estimates).}
\label{fig:turnover}
\end{figure}


\section{Case Study}
\label{sec:cases}
We walk through two decisions from the once-mode grid (\S\ref{subsec:constructor}), both on the window's most consequential name, Intel (\texttt{INTC}): the first fixes the analyst evidence and varies the constructor, the second fixes the constructor and varies the analyst capture.

\subsection{Same evidence, different constructor}
\label{case:headtohead}
Handed the \emph{same} frozen \texttt{gpt-5} analyst capture (seed~1, $K{=}50$), in which the analysts rank \texttt{INTC} second of the $50$ candidates, the strongest and weakest constructors diverge sharply (Table~\ref{tab:headtohead}). \texttt{gpt-5} \emph{re-weights} the candidate pool: it keeps $15$ of the $20$ same-pool names (Jaccard $0.75$), tilts toward conviction without breaching the $10\%$ cap, and holds \texttt{INTC} at $8.5\%$. \texttt{gpt-4o-mini} instead departs: it concentrates ten names, each pinned at the mandate cap (Jaccard $0.43$), and drops \texttt{INTC} altogether.

\textbf{One name drives the gap.} Both books are scored on identical prices, yet \texttt{gpt-5} returns $+11.5\%$ against \texttt{gpt-4o-mini}'s $+0.7\%$, and attribution explains nearly all of the difference: of \texttt{gpt-5}'s ${+}\$110$k attributed P\&L, \texttt{INTC} alone contributes ${+}\$104$k, so dropping the name its own analysts ranked second leaves \texttt{gpt-4o-mini} near breakeven (${+}\$3$k). The pattern is robust within this capture and candid about its own fragility: across the three seeds \texttt{gpt-5} always holds \texttt{INTC} and returns $11$--$13\%$, whereas \texttt{gpt-4o-mini} holds it in only one seed, and in exactly that seed it too reaches $+13.1\%$. Construction skill here is less about breadth than about \emph{acting on the strongest signal}, and a single trending name accounts for much of the spread between constructors (\S\ref{sec:limitations}).

\begin{table}[t]
\centering\footnotesize
\setlength{\tabcolsep}{5pt}
\begin{tabular}{@{}lrr@{}}
\toprule
 & \textbf{\texttt{gpt-5}} & \textbf{\texttt{gpt-4o-mini}} \\
\midrule
Net return              & $+11.5\%$ & $+0.7\%$ \\
Ann.\ Sharpe            & $2.71$    & $0.29$ \\
Holdings ($N$)          & $15$      & $10$ \\
Max name weight         & $9.0\%$   & $9.3\%$ \\
Overlap w/ EW (Jaccard) & $0.75$    & $0.43$ \\
\midrule
\rowcolor{grouprow}\multicolumn{3}{@{}l}{\textbf{\texttt{INTC} attribution}}\\
Held \texttt{INTC} & $8.5\%$ & dropped \\
\texttt{INTC} P\&L      & $+\$104$k & --- \\
Total P\&L              & $+\$110$k & $+\$3$k \\
\bottomrule
\end{tabular}
\caption{Two constructors on one byte-identical analyst capture (\texttt{gpt-5} analysts, seed~1, $K{=}50$). The stronger constructor re-weights the candidate pool and keeps the top-ranked name; the weaker one concentrates at the mandate cap and drops it. Single-cell values; three-seed means appear in Table~\ref{tab:main}. Source: \texttt{constructor\_eval} run artifacts.}
\label{tab:headtohead}
\end{table}

\subsection{Same constructor, different evidence}
\label{case:lever}
Holding the constructor fixed and swapping the analyst capture isolates the upstream effect. Because the liquidity gate is deterministic, both captures hand the constructor the same $50$-name pool, \texttt{INTC} included; only the analysts' read of it differs. The gpt-5 analysts converge bullish and confident (events $+0.9$ at confidence $0.9$, technical $+0.5$) and rank \texttt{INTC} second of fifty, whereas the DeepSeek analysts are divided (technical $-0.3$, lower event confidence) and rank it seventh. Handed the gpt-5 capture, the gpt-5 constructor buys \texttt{INTC} at $8.5\%$ and returns $+11.5\%$; handed the DeepSeek capture it holds none and returns $+3.1\%$, though only the evidence changed (Table~\ref{tab:lever}). The flip is not model-specific: the DeepSeek constructor moves the same way, and the weakest constructors hold \texttt{INTC} under neither capture, so the constructor can act only on what the analysts surface and upstream evidence quality sets the ceiling (\S\ref{subsec:analyst}).

\begin{table}[t]
\centering\footnotesize
\setlength{\tabcolsep}{5pt}
\begin{tabular}{@{}lrr@{}}
\toprule
 & \textbf{DeepSeek cap.} & \textbf{\texttt{gpt-5} cap.} \\
\midrule
\rowcolor{grouprow}\multicolumn{3}{@{}l}{\textbf{Analyst composite} (withheld from constructor)}\\
\texttt{INTC} composite rank   & $\#7/50$ & $\#2/50$ \\
\texttt{INTC} composite score  & $0.40$   & $0.47$ \\
\midrule
\rowcolor{grouprow}\multicolumn{3}{@{}l}{\textbf{\texttt{gpt-5} constructor}}\\
\quad \texttt{INTC} weight     & $0\%$    & $8.5\%$ \\
\quad Net return               & $+3.1\%$ & $+11.5\%$ \\
\addlinespace[2pt]
\rowcolor{grouprow}\multicolumn{3}{@{}l}{\textbf{\texttt{DeepSeek-V3.2} constructor}}\\
\quad \texttt{INTC} weight     & $0\%$    & $8.0\%$ \\
\quad Net return               & $+2.1\%$ & $+11.2\%$ \\
\bottomrule
\end{tabular}
\caption{The analyst-tier lever: one constructor held fixed while the analyst capture is swapped (seed~1, $K{=}50$). Both captures hand the constructor the same liquidity-gated pool, including \texttt{INTC}; the gpt-5 analysts rank it second and the DeepSeek analysts seventh, and the constructor holds it only under the higher-conviction capture. Single-cell values. Source: \texttt{constructor\_eval} run artifacts.}
\label{tab:lever}
\end{table}

\section{Reproducibility Details}
\label{sec:repro}
Every number we report is backed by a saved run directory. This appendix describes how those runs are produced and what each one stores.

\paragraph{Environment and access.} All runs use Python~3.11 with pandas~3.0 and read the single content-hashed freeze of \S\ref{sec:setup}. Model backbones are accessed under a bring-your-own-key scheme through one OpenRouter endpoint, with the key taken from the environment and never logged, and every call decodes at temperature~0. The analyst and constructor tiers are configured independently, so the two can share a backbone or differ.

\paragraph{Shared configuration.} Unless stated otherwise, both modes use the balanced mandate (long-only, a $10\%$ per-name cap, at most $20$ names, and no cash buffer), a liquidity-gated pool of $K{=}50$ candidates, and the deterministic critic and per-name liquidity clip on the critical path. The aggregated composite is withheld from the constructor and the candidate order is shuffled, the mandate is supplied to the constructor in its prompt, regime evidence uses the compact render of \S\ref{sec:allocator}, and the cross-section is PIT membership-gated. The six analysts run on \texttt{DeepSeek-V3.2} throughout, so only the constructor and, in once-mode, the analyst capture vary.

\paragraph{Once-mode grid.} The central experiment follows a capture-once, replay-many design. We run the six-analyst tier once for each (analyst backbone, seed) pair and freeze its output as an analyst capture, then replay that frozen capture through every constructor with no further analyst calls. Two analyst backbones (\texttt{DeepSeek-V3.2} and \texttt{gpt-5}) and three seeds give six captures, which we replay through five constructors (\texttt{gpt-5}, \texttt{claude-opus-4.7}, \texttt{DeepSeek-V3.2}, \texttt{qwen3-max}, and \texttt{gpt-4o-mini}) at $K\in\{30,40,50\}$, for the $90$ LLM cells and $216$ no-LLM baseline cells of a single decision date. Because the evidence is byte-identical across constructors, the constructor is the only component that changes. We also ran \texttt{gemini-2.5-flash} but exclude it from the results: it returned an essentially all-cash book on every input, including bullish controls, so its weights are not a meaningful allocation.

\paragraph{Daily sweep.} The daily experiment runs the full pipeline with no replay, once per constructor, rebalancing on each of the 44 trading days at $K{=}50$ with the same \texttt{DeepSeek-V3.2} analysts. Each run invokes the \texttt{agents.portfolio} module with the \texttt{llm\_tiered} provider, the balanced risk preset, and a daily rebalance over the 2026-03-02 to 2026-05-01 window, changing only the constructor model between runs; the resolved settings are recorded in the run manifest. Rebalancing is held to the first bar at least 60 minutes after the open and then to once per day, both fixed canonical behavior. As these are single runs, we read them as point estimates.

\paragraph{Run artifacts.} Each run writes a self-describing directory. A manifest records the full resolved configuration together with the dataset and capture hashes and token usage; the per-stage logs hold the analyst score matrix, the composite, and the constructor's proposed and executed weights; the trade and portfolio logs store the quoted bid, ask, and mid behind every fill alongside the reconciled equity curve; the evaluation logs hold the full metric set with the constraint-adherence report, the contamination certificate, and the backtest-bias checklist; and a redacted log preserves the prompts and replies. The once-mode grid and the daily sweep are kept under their own experiment directories, so any reported value can be traced to the run that produced it.

\end{document}